\documentclass[runningheads]{llncs}
\usepackage[T1]{fontenc}
\usepackage{graphicx,verbatim}
\usepackage{color}
\usepackage{url}

\usepackage{booktabs}
\usepackage{amssymb}
\usepackage{subfig}
\begin{document}
	\title{BreastMammo and DenseMammo: Benchmarks for Mammography Domain Generalization}
	\titlerunning{BreastMammo and DenseMammo}
	%
	\author{Hongyi Pan\inst{1} \and
		Gorkem Durak\inst{1} \and
		Halil Ertugrul Aktas\inst{1} \and
		Andrea Mia Bejar\inst{1} \and
		Mustafa Ege Seker\inst{2} \and
		Nebile Alibeyoglu\inst{3} \and
		Rumeysa Guclu\inst{3} \and
		Rana Gunoz Comert Bozkurt\inst{3} \and
		Sibel Ozkan Gurdal\inst{4} \and
		Neslihan Cabioglu\inst{3} \and
		Beyza Ozcinar\inst{3} \and
		Ravza Yilmaz\inst{3} \and
		Vahit Ozmen\inst{5} \and
		Erkin Aribal\inst{6} \and
		Sukru Mehmet Erturk\inst{3} \and
		Yalda Zafari\inst{7}\and
		Mohamed Mabrok\inst{7}\and
		Kayhan Batmanghelich\inst{8}\and
		Mohammad Yaqub\inst{9}\and
		Ziyue Xu\inst{10} \and
		Ulas Bagci\inst{1}}
	\authorrunning{H. Pan et al.}
	\institute{Northwestern University\\
		\email{\{hongyi.pan, ulas.bagci\}@northwestern.edu} \and
		University of Wisconsin-Madison \and
		Istanbul University \and
		Namik Kemal University \and
		Istanbul Florence Nightingale Hospital \and
		Acibadem Mehmet Ali Aydinlar University \and
		Qatar University\and
		Boston University \and
		Mohamed bin Zayed University of Artificial Intelligence\and
		NVIDIA}
	
	
	
	\maketitle              
	\begin{abstract}
		Breast density classification is a critical component of breast cancer risk assessment, yet AI models often struggle to generalize across clinical sites due to vendor-specific acquisition styles. In this work, we introduce two new datasets, \textit{BreastMammo} and \textit{DenseMammo}, to facilitate robust multi-view mammography research. We propose a domain generalization framework that utilizes a foreground-only histogram matching protocol to resolve the domain shift issue arising from disparate clinical sources. Internal evaluation using a 5-fold cross-validation protocol demonstrates the efficacy of our approach, with the Swin Transformer backbone achieving a peak AUC of 98.32\% for density classification. External evaluation on the TNMammo and LUMINA datasets demonstrates that the proposed approach consistently reduces domain shift, significantly outperforming prominent domain generalization paradigms, including MixStyle and Discrete-Fourier-Transform-based frameworks.
		
		
	\end{abstract}
	\section{Introduction}
	
	Breast cancer remains the most prevalent malignancy and a leading cause of cancer-related mortality among women worldwide~\cite{ries2008seer,lukasiewicz2021breast,wilkinson2022understanding}. 
	A critical factor in both risk assessment and diagnostic accuracy is breast density. High breast density (types C and D) is not only an independent risk factor for developing cancer but also creates a ``masking effect'' that significantly reduces the sensitivity of mammography~\cite{mann2022breast}, as dense fibroglandular tissue can obscure underlying lesions. Consequently, accurate and automated breast density classification is paramount for clinical decision-making, determining which patients require supplemental screening methods such as MRI or ultrasound.
	
	Despite the potential of deep learning to automate this classification, the development of robust models is hindered by two major challenges: the scarcity of large-scale, multi-center datasets and the pervasive issue of domain shift. In full-field digital mammography (FFDM), domain shifts are driven by variations in hardware acquisition parameters and proprietary vendor-specific post-processing algorithms. These factors create energy-driven appearance shifts that alter pixel-intensity distributions. When models are trained on merged data from multiple institutions without proper alignment, these disparate vendor styles can introduce contradictory signals where the inclusion of more data actually degrades performance compared to training on a single site alone.
	
	In this work, we address these challenges by proposing two new FFDM datasets alongside a tissue-specific domain generalization framework. Specifically, our contributions are summarized as the following: First, we introduce \textit{BreastMammo}, a diagnostic two-view dataset containing 894 images from 447 pathology-confirmed cases, and \textit{DenseMammo}, a screening four-view dataset comprising 2,480 images from 620 patients. They both contain expert-annotated breast density labels. Second, we leverage these resources to establish a rigorous, standardized multi-view benchmark for automated breast density classification to help manage the masking effect in dense tissue. Third, we introduce a domain generalization method based on foreground-only histogram matching to resolve the domain shift issue. It significantly improved classification performance on unseen clinical domains such as TNMammo~\cite{nguyen2025tn} and LUMINA~\cite{pan2026lumina}.
	
	\section{Related Works}
	\noindent\textbf{Mammography Dataset and Benchmarking:} The development of computer-aided diagnosis (CAD) systems for breast cancer has historically relied on digitized film mammography datasets such as MIAS~\cite{suckling1994mammographic} and CBIS-DDSM~\cite{lee2017curated}. While foundational, these resources have paved the way for modern FFDM applications. Recent efforts have introduced large-scale FFDM datasets, including INbreast~\cite{moreira2012inbreast}, KAU-BCMD~\cite{alsolami2021king}, RSNA~\cite{carr2022rsna}, CMMD~\cite{cai2023online}, CDD-CESM~\cite{khaled2022categorized}, VinDR-Mammo~\cite{nguyen2023vindr}, and LUMINA~\cite{pan2026lumina}, providing high-resolution images with a variety of clinical labels. 
	Building upon this progress, there is an ongoing need for multi-view benchmarks dedicated to breast density classification--a task that is inherently sensitive to intensity variations and vendor-specific post-processing. 
	
	\noindent\textbf{Domain Generalization:} 
	Domain generalization (DG) has become a crucial area of research for ensuring the clinical reliability of mammography models across different institutions. To learn domain-invariant representations, prominent feature-space style augmentation methods, such as MixStyle~\cite{zhou2021domain}, dynamically blend the feature map statistics (mean and variance) of different instances within a mini-batch during the training forward pass. While MixStyle effectively introduces latent style perturbations to force feature regularization, feature-level manipulation lacks explicit anatomical grounding and can inadvertently alter or corrupt subtle, localized texture patterns that are critical for accurate density stratification. 
	Alternatively, frequency-domain frameworks utilize the Discrete Fourier Transform (DFT) to achieve generalization by interpolating in a continuous frequency space~\cite{xu2021fourier,liu2021feddg,pan2024domain,pan2025frequency}, primarily swapping or perturbing the amplitude spectrum to capture low-level acquisition styles. However, these DFT-based techniques are fundamentally limited in mammography because they must operate on a fixed rectangular window, which inadvertently incorporates the large, uninformative black background into the style alignment. 
	
	\section{BreastMammo and DenseMammo Datasets}
	\textit{BreastMammo} is a two-view diagnostic dataset comprising 894 images from 447 patients. Each patient profile includes paired Craniocaudal (CC) and Mediolateral Oblique (MLO) views of a single breast. The cohort consists of 576 benign and 318 malignant images with pathology-confirmed labels, BI-RADS assessment scores, and breast density categories (ACR categories A--D). \textit{DenseMammo} is a large-scale four-view screening dataset for breast density classification. It contains 2,480 images from 620 patients. Adhering to standard clinical protocols, each patient provided a full screening set consisting of four views: Left CC, Left MLO, Right CC, and Right MLO. All images are provided with expert-annotated breast density labels. 
	
\begin{figure}[tb]
		\centering
		\subfloat[\textit{BreastMammo}.\label{fig:breastmammo_combined}]{
			\includegraphics[width=.6\linewidth]{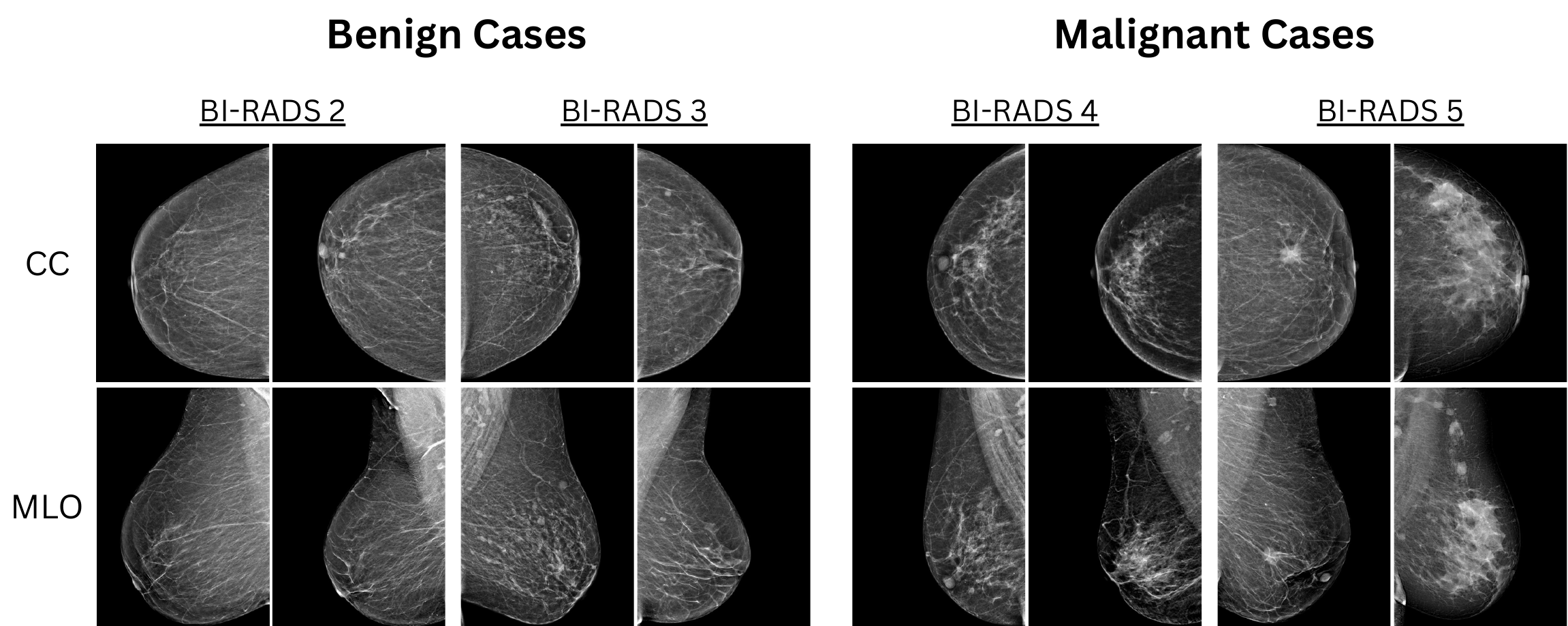}\hfill
			\includegraphics[width=.3\linewidth]{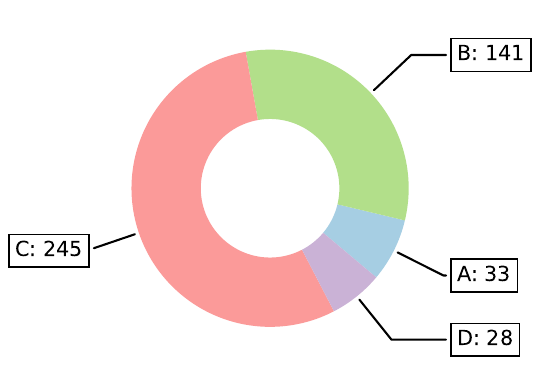}
		}
		\vspace{-10pt}\\
		\subfloat[\textit{DenseMammo}.\label{fig:densemammo_combined}]{
			\includegraphics[width=.6\linewidth]{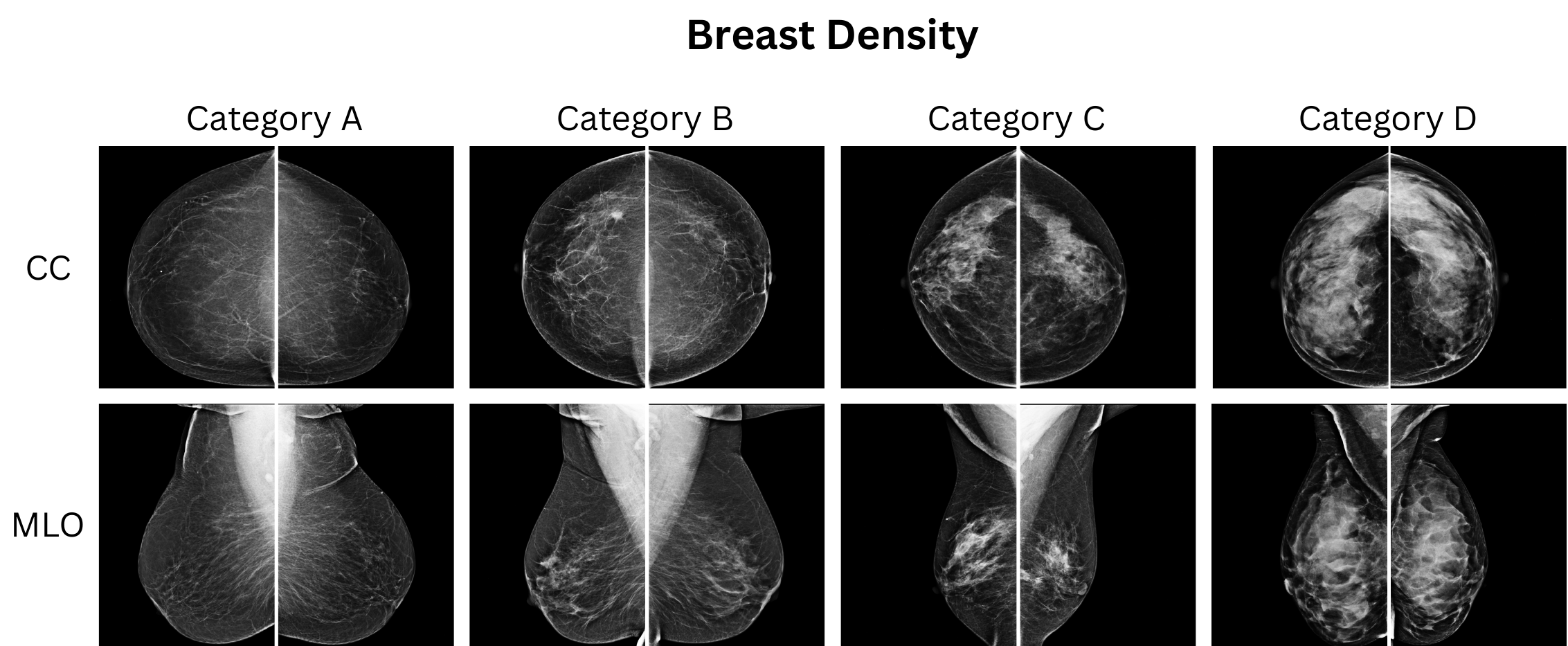}\hfill
			\includegraphics[width=.3\linewidth]{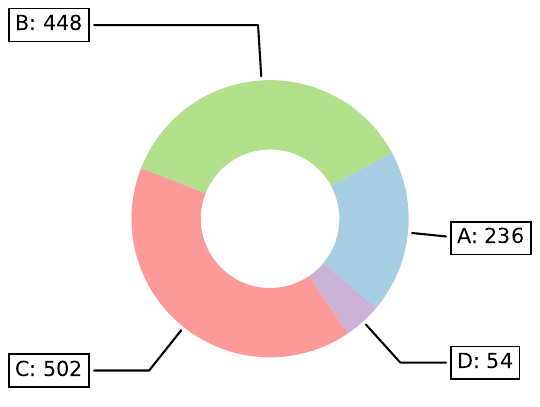}
		}\vspace{-8pt}
		\caption{\textbf{Representative Samples And Density Distributions.}}\vspace{-15pt}
		\label{fig:samples}
	\end{figure}

	All imaging data in the two datasets were fully anonymized prior to algorithmic processing. These datasets are released in both DICOM and 16-bit PNG formats to facilitate reproducible research in the community. Their samples and density distributions are visualized in Fig.~\ref{fig:samples}. This study received approval from the Institutional Review Board (IRB) of the participating institutions (Approval Date: 11/20/2024; protocol number: 3026880). Due to the study's retrospective nature, informed consent was waived. All images were collected from the hospital’s Picture Archiving and Communication System (PACS) under the guidance of three experienced breast radiologists, R.G.C.B., R.Y., and E.A., who have 10, 21, and 35 years of experience, respectively.

	\begin{figure}[tb]
		\centerline{\includegraphics[width=.9\linewidth]{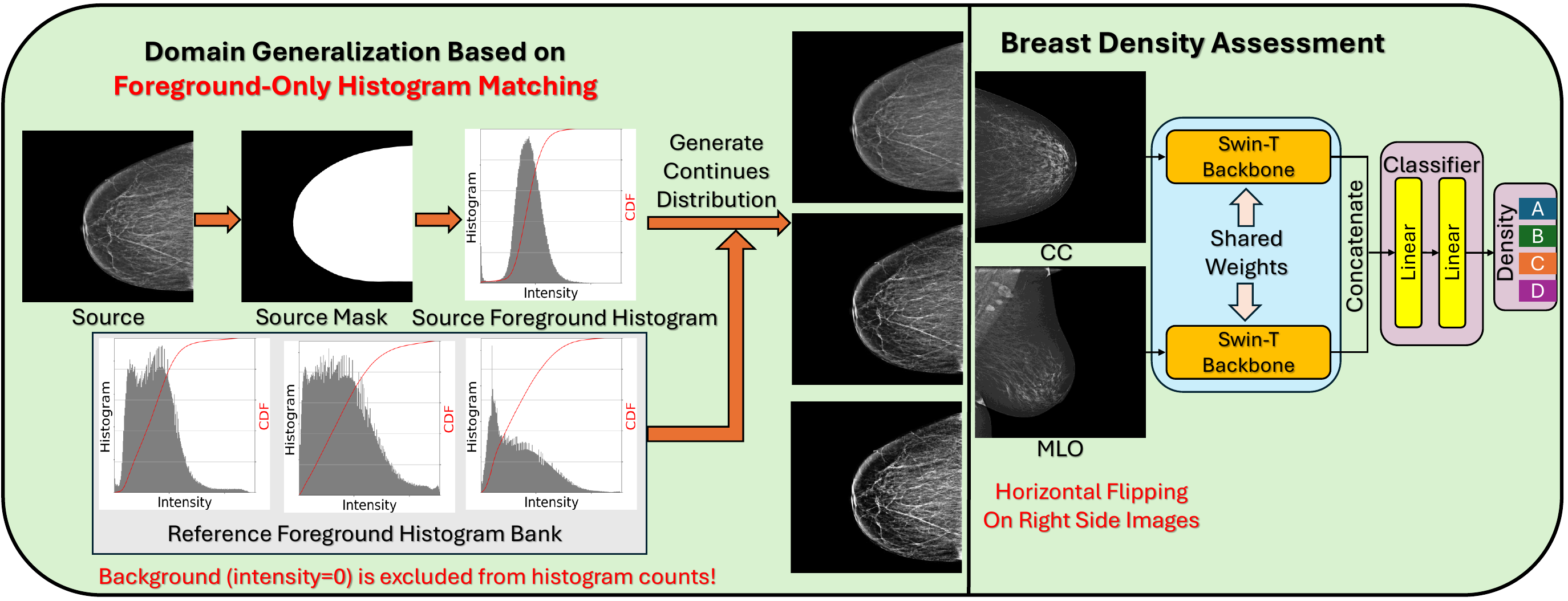}}\vspace{-8pt}
		\caption{\textbf{Multi-View Domain Generalization And Classification Pipeline.} Our domain generalization method integrates foreground-only histogram matching to generate a continuous distribution of styles.}
		\label{fig:pipeline}
	\end{figure}
	
	\begin{figure}[tb]
		\centering
		\begin{minipage}{0.162\linewidth}
			\centering
			\includegraphics[width=\linewidth, height=\linewidth]{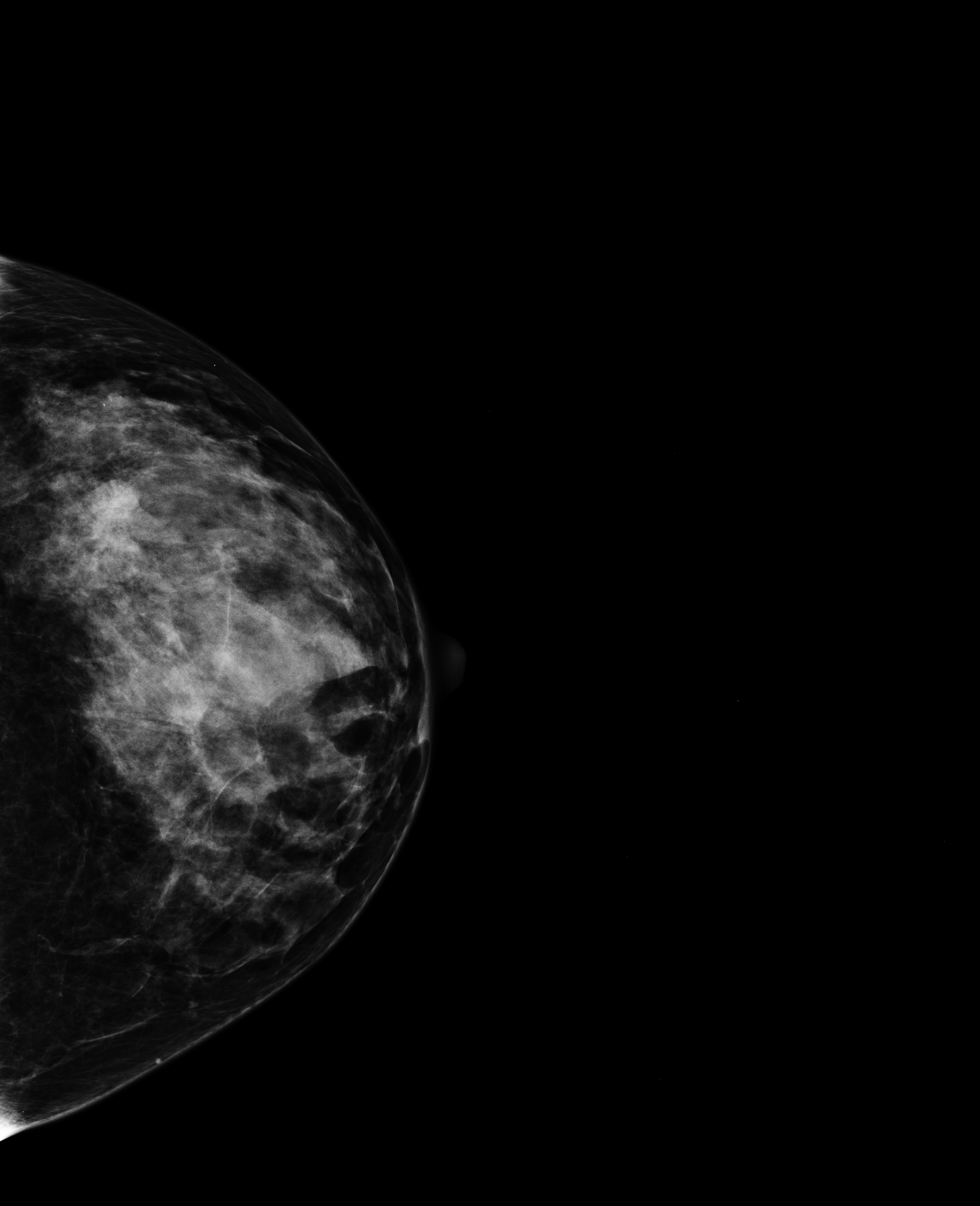}
			\subfloat[Source.]{\includegraphics[width=\linewidth, height=\linewidth]{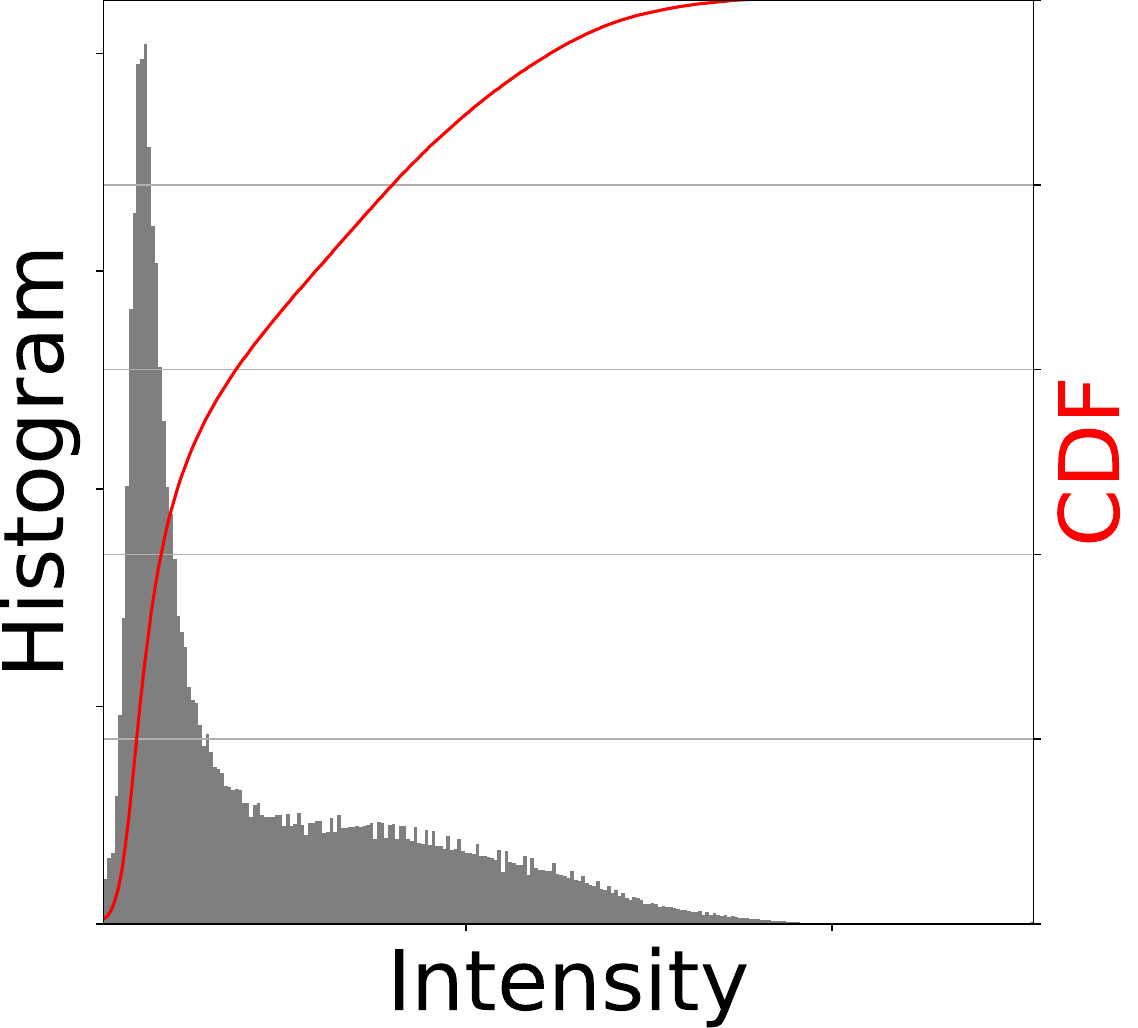}\label{fig:match_source}}
		\end{minipage}\hfill
		\begin{minipage}{0.162\linewidth}
			\centering
			\includegraphics[width=\linewidth, height=\linewidth]{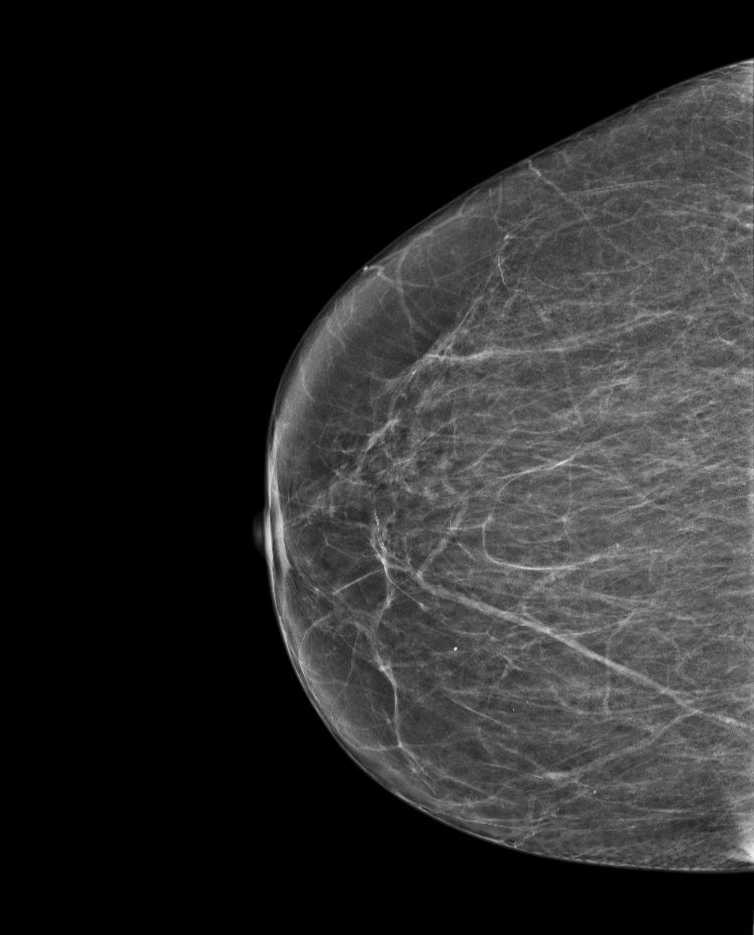}
			\subfloat[Reference.]{\includegraphics[width=\linewidth, height=\linewidth]{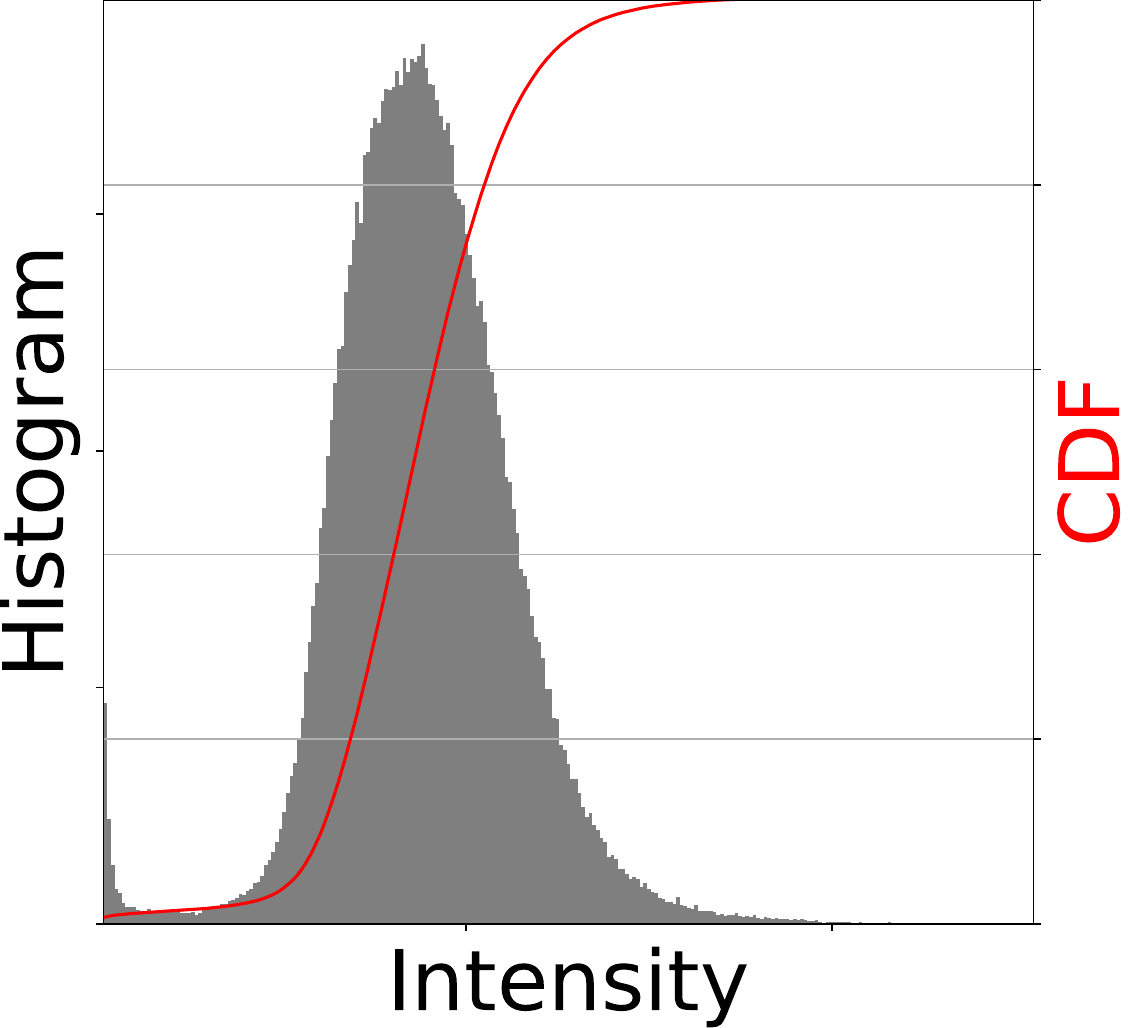}\label{fig:match_ref}}
		\end{minipage}\hfill
		\begin{minipage}{0.162\linewidth}
			\centering
			\includegraphics[width=\linewidth, height=\linewidth]{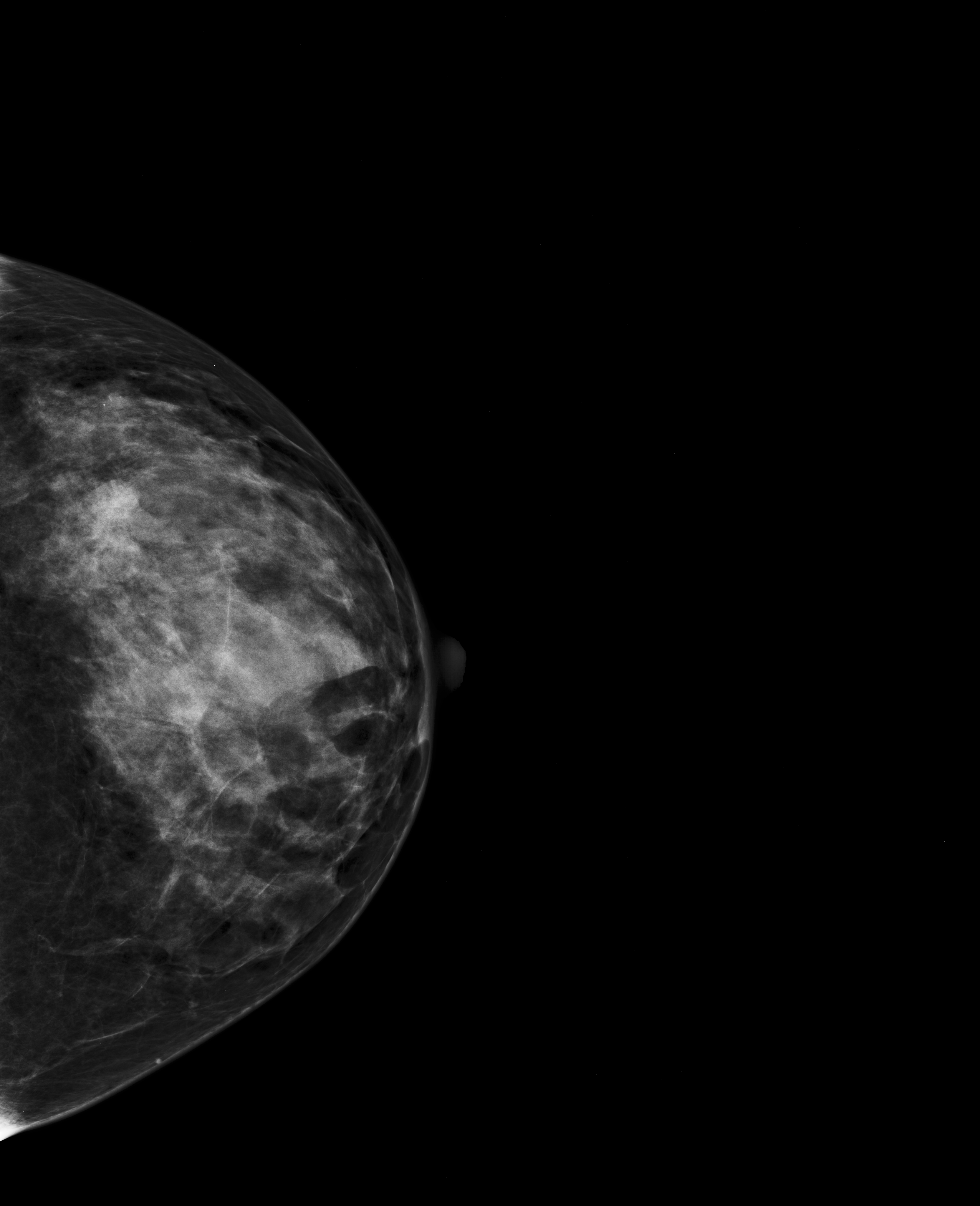}
			\subfloat[$\alpha = 0.25$.]{\includegraphics[width=\linewidth, height=\linewidth]{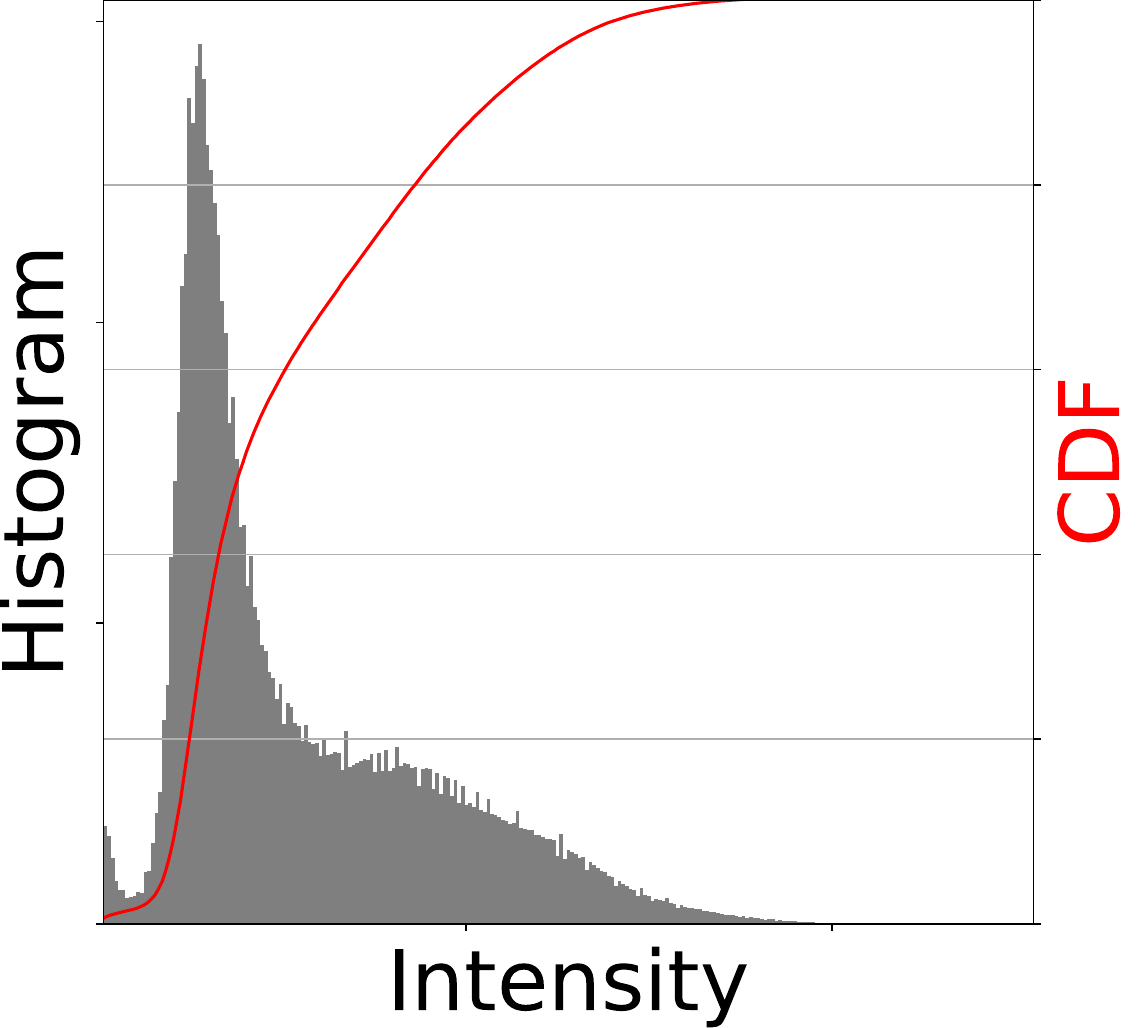}\label{fig:match_025}}
		\end{minipage}\hfill
		\begin{minipage}{0.162\linewidth}
			\centering
			\includegraphics[width=\linewidth, height=\linewidth]{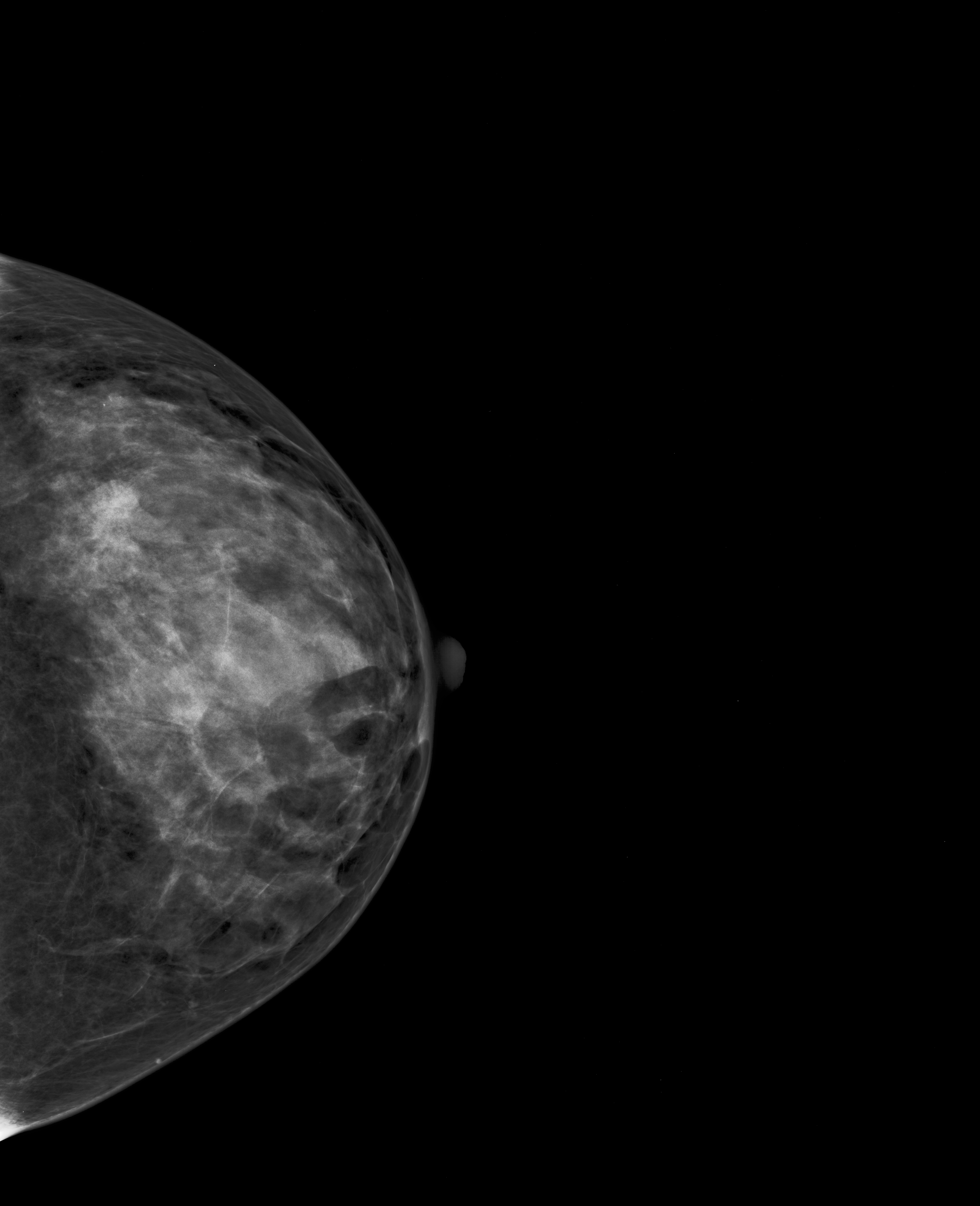}
			\subfloat[$\alpha = 0.5$.]{\includegraphics[width=\linewidth, height=\linewidth]{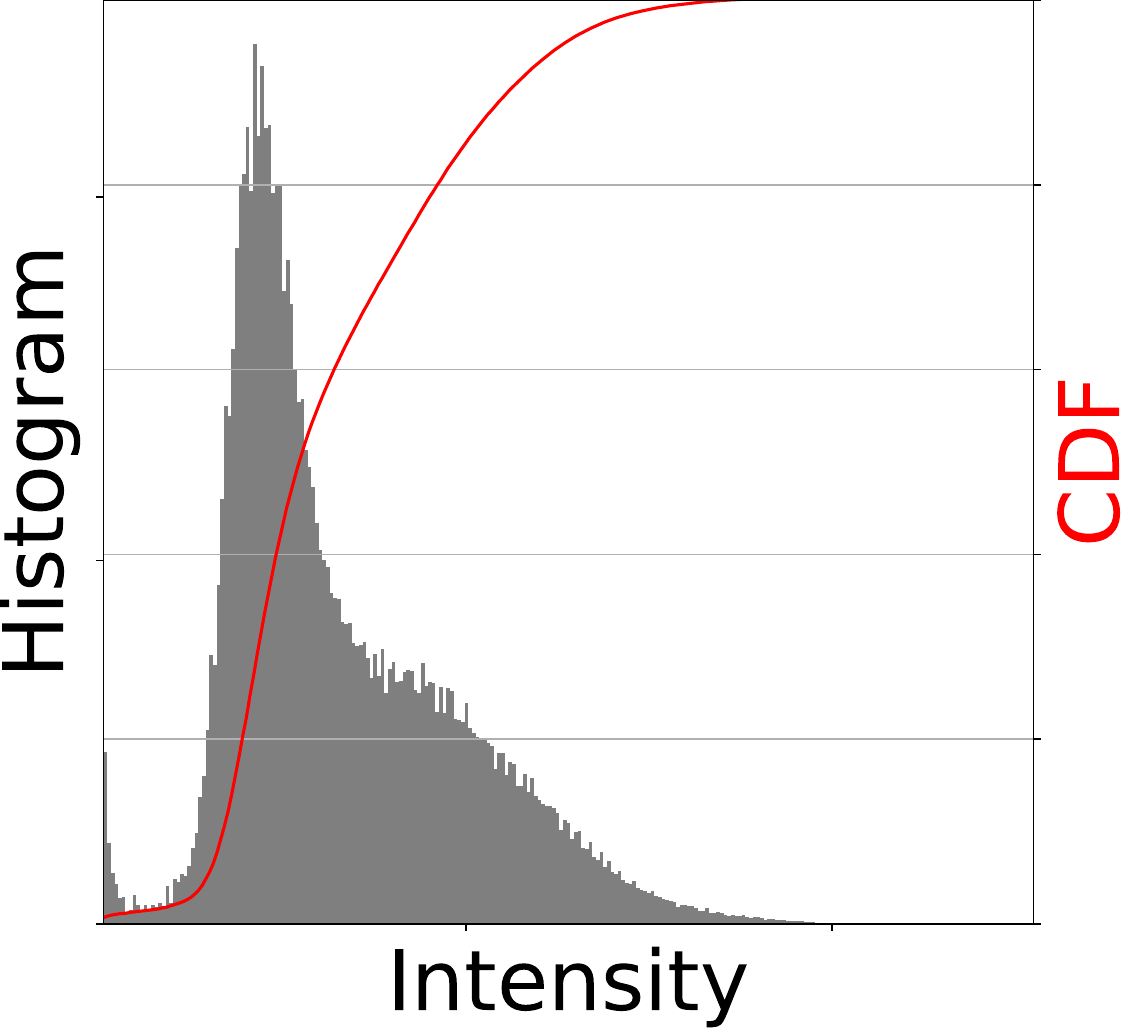}\label{fig:match_050}}
		\end{minipage}\hfill
		\begin{minipage}{0.162\linewidth}
			\centering
			\includegraphics[width=\linewidth, height=\linewidth]{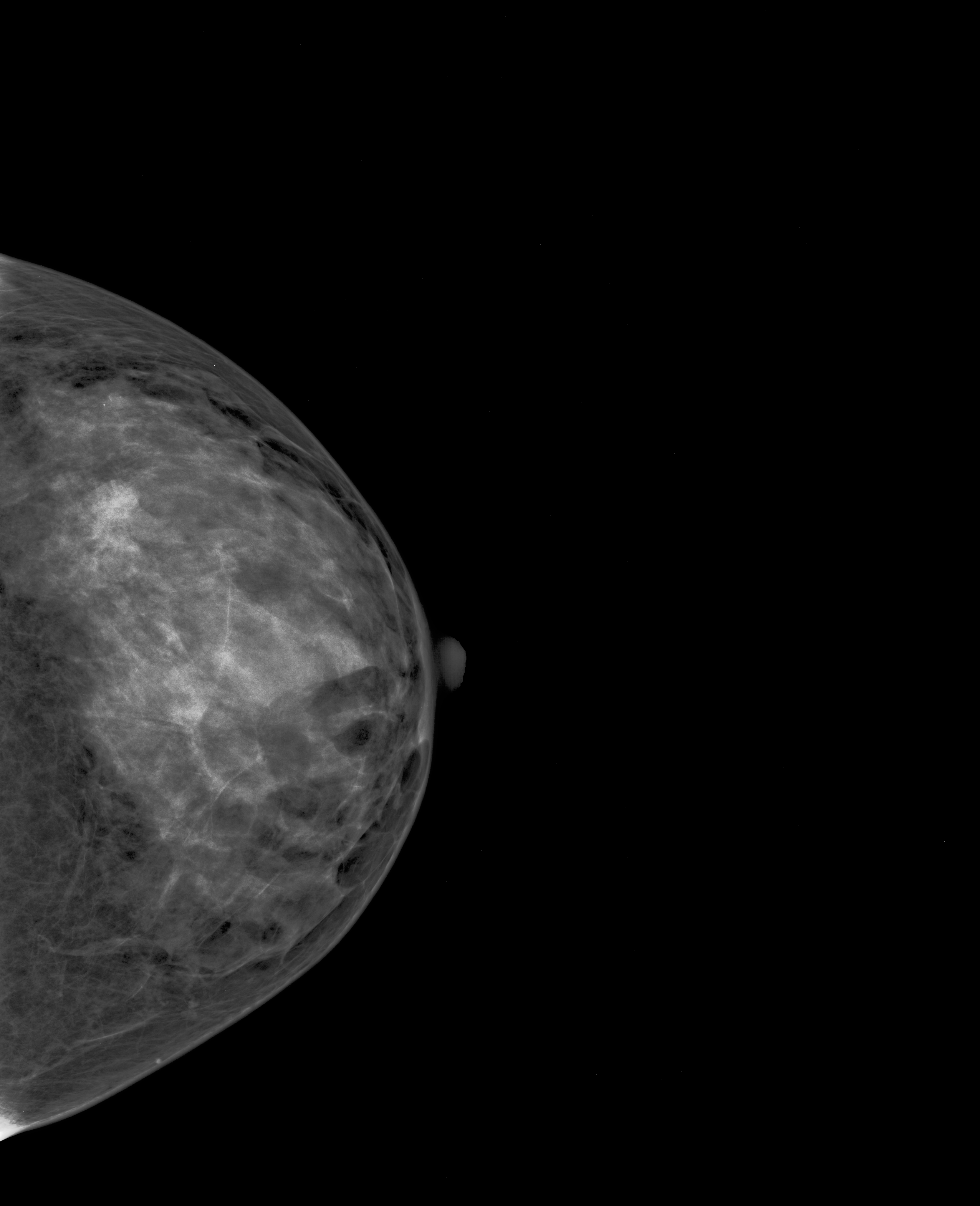}
			\subfloat[$\alpha = 0.75$.]{\includegraphics[width=\linewidth, height=\linewidth]{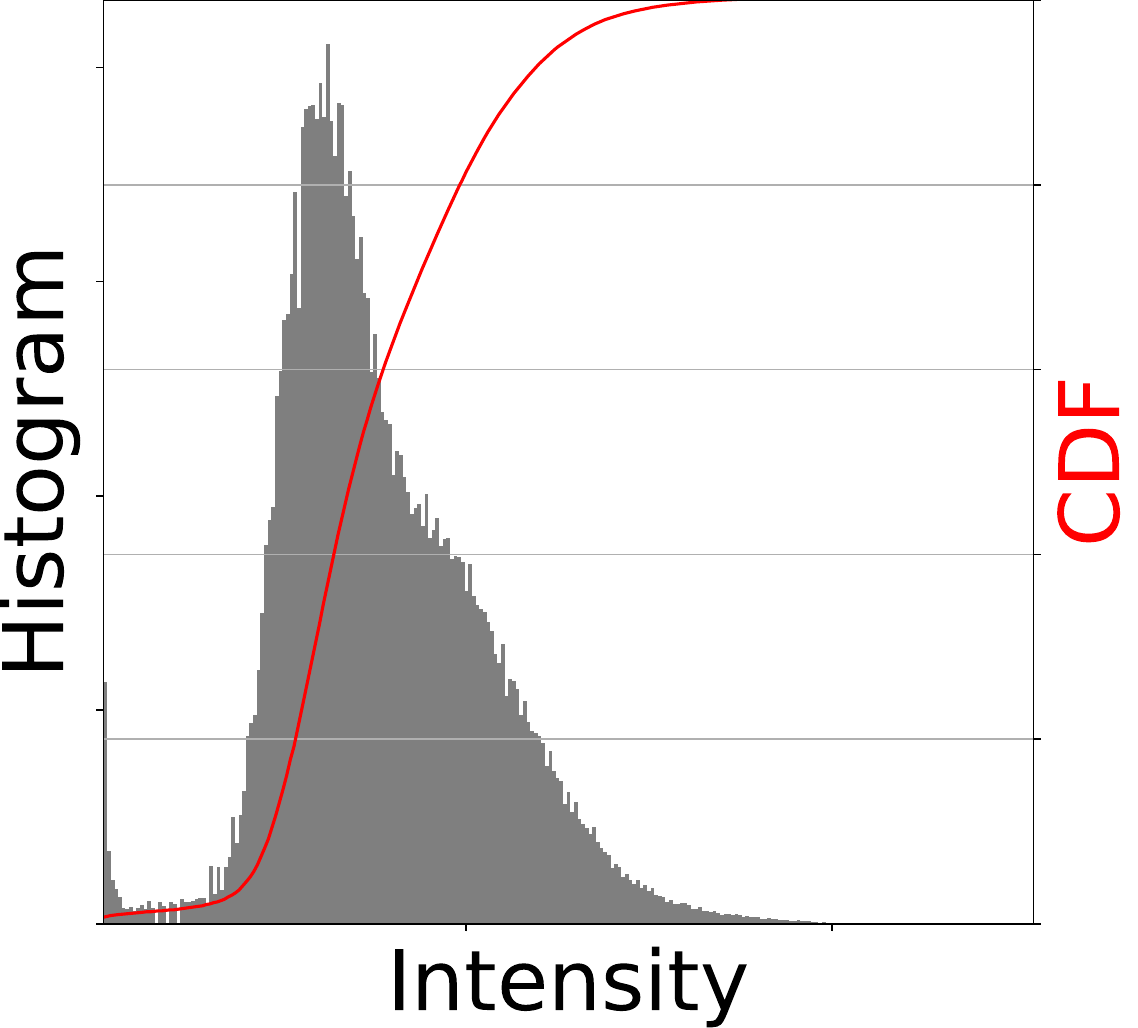}\label{fig:match_075}}
		\end{minipage}\hfill
		\begin{minipage}{0.162\linewidth}
			\centering
			\includegraphics[width=\linewidth, height=\linewidth]{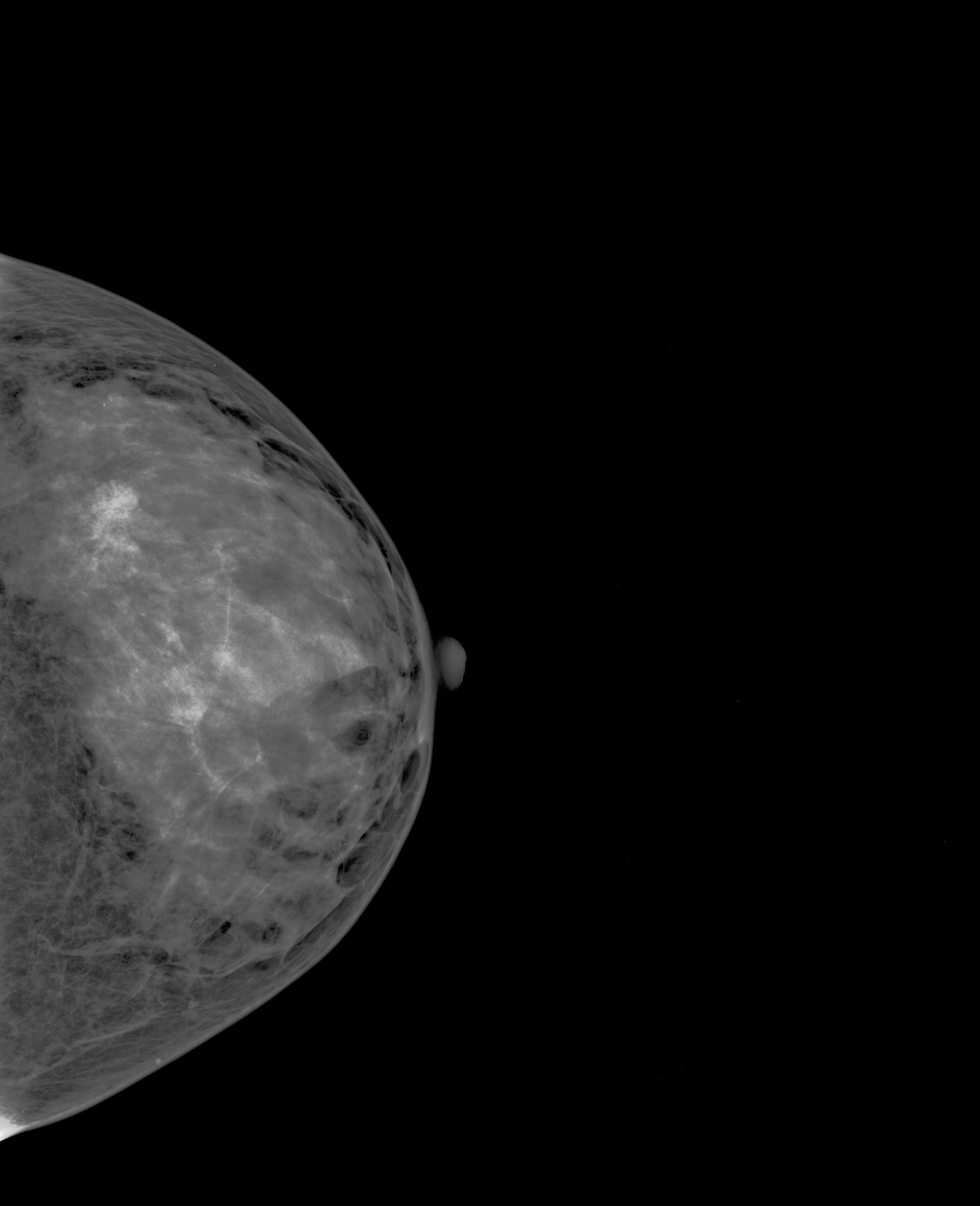}
			\subfloat[$\alpha = 1$.]{\includegraphics[width=\linewidth, height=\linewidth]{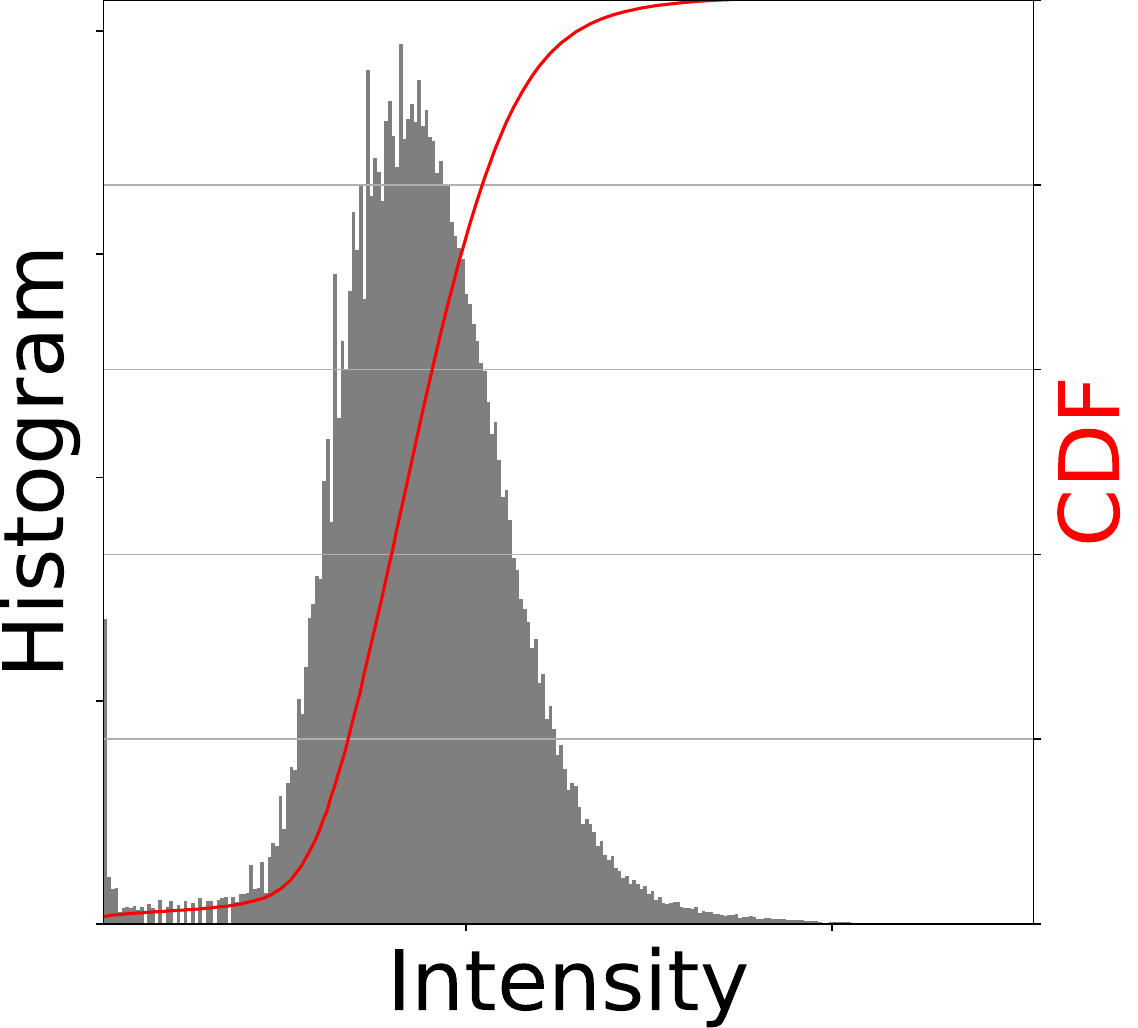}\label{fig:match_100}}
		\end{minipage}
		\vspace{-5pt}
		\caption{\textbf{Intensity Alignment via Foreground-Only Histogram Matching.} 
			Top: mammograms. Bottom: Intensity histograms and CDFs. The objective is to adapt the acquisition style of the source image (a) to match the reference target (b). The progressive transition (c--f) demonstrates the synthetic variations. }\vspace{-15pt}
	\label{fig: matching}
\end{figure}

\section{Domain Generalization and Density Classification}
As shown in Fig.~\ref{fig:pipeline}, our pipeline consists of two main modules: a foreground-only histogram matching based domain generalization and a Swin Transformer~\cite{liu2021swin} based two-view classification network.

\subsection{Domain Generalization via Histogram Matching}
Our domain generalization approach builds upon the histogram-based data harmonization framework introduced by LUMINA~\cite{pan2026lumina}, which demonstrated the efficacy of foreground-only pixel-space alignment in mitigating vendor-induced intensity drift. While LUMINA utilizes histogram to map images to a specific low-energy reference target for static harmonization, we extend this principle to a target-free domain generalization framework that dynamically bridges domain gaps by decoupling breast tissue style from its morphological content during training. Following traditional histogram matching protocols~\cite{rolland2000fast,shen2007image,tu2013histogram,shapira2013multiple}, given a source image $\mathbf{I}_s$ from domain $\mathbb{D}_s$ and a reference image $\mathbf{I}_r$ from domain $\mathbb{D}_r$, an intensity mapping function $\mathcal{T}$ is established by aligning their cumulative distribution functions (CDFs). Crucially, because mammograms contain large, non-clinical black background regions that bias statistical counts and degrade matching quality, these uninformative regions are explicitly excluded by strictly evaluating pixels with an intensity value, $p$, greater than zero ($p > 0$). The histograms and their normalized CDFs ($\bar{C}_s$ and $\bar{C}_r$) are subsequently computed solely over these isolated breast tissue foreground masks, mapping the source pixel intensities such that $\bar{C}_s(p) \approx \bar{C}_r(\mathcal{T}(p))$, where $\mathcal{T}(\cdot)$ denotes the intensity mapping function. Moreover, we follow~\cite{liu2021feddg} to introduce a hyperparameter $\alpha \in [0, 1]$ to control the intensity of the style shift for the final synthetic image $\mathbf{I}_o$ by linearly blending the matched result with the original source characteristics:
\begin{equation}
	\mathbf{I}_o(x, y) = \alpha \cdot \mathcal{T}(\mathbf{I}_s(x, y)) + (1-\alpha) \cdot \mathbf{I}_s(x, y), \label{eq: histogram mapping final}
\end{equation}
where $(x, y)$ denote the spatial coordinates.
By varying $\alpha$, we generate a series of synthetic samples that balance the original source characteristics with the target domain's intensity profile. As illustrated in Fig.~\ref{fig: matching}, this protocol effectively normalizes vendor-specific contrast variations while preserving the underlying anatomical structures necessary for diagnosis. During training, our pairing mechanism operates under two strict clinical constraints to ensure data consistency. First, for each training instance originating from a source domain, a reference image is dynamically and randomly sampled from the alternative target domain's training pool to maximize cross-institutional style exposure. Second, because clinical mammography relies on multi-view examinations, we enforce inter-view style synchronization that the identical reference style target is utilized across all corresponding views (CC and MLO) for a given patient case. 
For each raw patient case, we generate four synthetic variations spanning progressive style strengths ($\alpha \in \{0.25, 0.5, 0.75, 1\}$) to establish a continuous training manifold.

\subsection{Multi-View Classification Architecture}
The aligned images are processed through a Swin-T backbone~\cite{liu2021swin}. For multi-view analysis, the pipeline implements shared weights across CC and MLO views. To ensure spatial consistency across bilateral images, we followed MammoClean~\cite{zafari2026mammoclean} to horizontally flip all right-side images before feature extraction. The hierarchical shifted-window attention of the Swin-T allows the model to capture fine pathological textures while maintaining a global context of the breast structure. Following feature extraction, the latent representations from each view are concatenated into a single feature vector. This unified representation is passed through a classification head consisting of linear layers to produce predictions for our clinical tasks: (1) density assessment, where models categorize breasts from \textit{BreastMammo} and \textit{DenseMammo} into one ACR density category (A, B, C, or D); and (2) pathology diagnosis, where models distinguish \textit{BreastMammo}'s breasts between benign and malignant cases based on the fused two-view features. We also provide the single-view diagnosis results, where we fine-tuned vanilla Swin-T and other models using standard transfer learning.

\section{Experimental Results}

\subsection{Experimental Setup}
All models were implemented using the PyTorch framework~\cite{paszke2019pytorch}. The training was conducted on a high-performance server equipped with 8 NVIDIA RTX A6000 GPUs. Each individual experiment was executed on a single GPU. Models were trained for 100 epochs with a batch size of 32 by the AdamW optimizer~\cite{loshchilov2017decoupled} with an initial learning rate of 0.0001. The training process followed a step learning rate scheduler that reduced the learning rate by a factor of 0.1 every 30 epochs. All images are rescaled to $224^2$ before being fed into the networks, and internal evaluation benchmarks also contain a higher resolution of $512^2$. 

For internal evaluation, we benchmarked several state-of-the-art backbones, including EfficientNet-B0~\cite{tan2019efficientnet}, DenseNet-121~\cite{huang2017densely}, ResNet-50~\cite{he2016deep}, and Swin-T~\cite{liu2021swin}. All backbones were pretrained on the ImageNet-1K dataset~\cite{deng2009imagenet} under $224^2$ resolution. Five-fold cross validation was adopted to establish a standard baseline benchmark. The internal evaluation was conducted using standard training protocols without any domain generalization techniques. 

External evaluation was conducted to evaluate our proposed histogram-based domain generalization in mitigating multi-center domain shifts using unseen datasets TNMammo~\cite{nguyen2025tn} and LUMINA~\cite{pan2026lumina}. Swin-T~\cite{liu2021swin} was selected as the backbone due to its superior internal performance. MixStyle~\cite{zhou2021domain} (feature-level augmentation) and DFT-based domain generalization~\cite{xu2021fourier,liu2021feddg} (low-frequency amplitude spectrum components swapping between source and reference samples) were implemented for comparison.
Crucially, because vanilla Fourier swapping introduces severe uninformative noise across the background, we implemented an optimized version that enforces a post-processing spatial mask to reset background pixels to zero where the original source image is zero.

\subsection{Experimental Results and Discussions}

\noindent\textbf{Internal evaluation:} The internal evaluation results in Table~\ref{tab:internal} demonstrate that the Swin-T architecture generally provides the most robust baseline. In the density classification task, Swin-T consistently outperformed CNN-based methods in AUC (96.53\% on \textit{BreastMammo} and 98.32\% on \textit{DenseMammo}). Notably, utilizing a larger input resolution ($512^2$) did not yield significant performance gains. This marginal improvement is likely attributed to a resolution mismatch, as the foundational backbones were initialized with weights pretrained on the ImageNet-1K dataset at a native $224^2$ resolution. 

\begin{table}[tb]
	\caption{\textbf{Internal Evaluation.} Results are presented as mean$\pm$std.}
	\centering
	\setlength{\tabcolsep}{1pt}
	\scriptsize
	\begin{tabular}{lcccccc}
		\toprule
		\textbf{Model} & \multicolumn{3}{c}{\textbf{Input Size: $224^2$}} & \multicolumn{3}{c}{\textbf{Input Size: $512^2$}} \\
		\cmidrule(lr){2-4} \cmidrule(lr){5-7}
		& \textbf{ACC(\%)} & \textbf{AUC(\%)} & \textbf{F1(\%)} & \textbf{ACC(\%)} & \textbf{AUC(\%)} & \textbf{F1(\%)} \\
		\midrule
		\multicolumn{7}{l}{\textbf{BreastMammo single-view diagnosis}}\\    \midrule
		EfficientNet-B0 & 89.04$\pm$10.93 & 90.79$\pm$12.52 & 78.62$\pm$25.99 & 87.81$\pm$10.75 & 91.69$\pm$10.54 & 76.88$\pm$26.23 \\
		DenseNet-121    & 89.94$\pm$11.07 & 92.12$\pm$10.89 & 80.43$\pm$24.98 & 86.80$\pm$8.92  & 93.27$\pm$9.12  & 78.88$\pm$17.56 \\
		ResNet-50       & 89.49$\pm$10.61 & 92.17$\pm$9.70  & 80.42$\pm$22.71 & 89.26$\pm$10.91 & 91.18$\pm$11.73 & 78.89$\pm$25.40 \\
		Swin-T          & \textbf{90.38$\pm$9.80} & \textbf{94.33$\pm$7.35} & \textbf{82.41$\pm$21.52} & 89.60$\pm$9.82 & 94.22$\pm$8.04 & 81.12$\pm$20.79 \\
		\midrule
		\multicolumn{7}{l}{\textbf{BreastMammo two-view diagnosis}}\\    \midrule
		EfficientNet-B0 & 89.54$\pm$11.85 & 92.67$\pm$9.42  & 80.38$\pm$26.36 & 87.52$\pm$12.40 & 93.83$\pm$6.39  & 75.76$\pm$30.75 \\
		DenseNet-121    & 89.09$\pm$12.18 & 93.01$\pm$9.12  & 77.08$\pm$31.04 & 84.59$\pm$9.16  & 93.83$\pm$8.75  & 74.68$\pm$21.15 \\
		ResNet-50       & \textbf{90.65$\pm$11.89} & \textbf{93.98$\pm$9.04} & \textbf{80.91$\pm$28.03} & 83.23$\pm$13.78 & 92.16$\pm$10.33 & 61.77$\pm$38.34 \\
		Swin-T          & 90.43$\pm$12.81 & 93.43$\pm$10.17 & 78.29$\pm$33.76 & 89.98$\pm$12.25 & 93.85$\pm$9.32  & 78.61$\pm$31.42 \\
		\midrule
		\multicolumn{7}{l}{\textbf{BreastMammo density classification}}\\    \midrule
		EfficientNet-B0 & 83.01$\pm$4.21  & 95.28$\pm$1.51  & 72.62$\pm$8.05  & 80.53$\pm$2.09  & 94.85$\pm$1.13  & 69.38$\pm$11.50 \\
		DenseNet-121    & 79.67$\pm$5.67  & 95.05$\pm$1.94  & 67.20$\pm$5.51  & 80.08$\pm$2.25  & 94.75$\pm$1.51  & 70.51$\pm$8.94  \\
		ResNet-50       & 77.87$\pm$3.56  & 92.90$\pm$2.75  & 62.87$\pm$13.49 & 76.52$\pm$3.05  & 92.86$\pm$2.32  & 60.95$\pm$10.27 \\
		Swin-T          & 82.34$\pm$4.29  & 95.49$\pm$1.43  & 67.27$\pm$10.40 & \textbf{84.77$\pm$3.91} & \textbf{96.53$\pm$0.53} & \textbf{75.89$\pm$12.04} \\
		\midrule
		\multicolumn{7}{l}{\textbf{DenseMammo density classification}}\\    \midrule
		EfficientNet-B0 & 85.40$\pm$2.76  & 97.70$\pm$0.66  & 74.70$\pm$7.70  & 87.82$\pm$2.77  & 97.92$\pm$0.70  & 81.90$\pm$7.11  \\
		DenseNet-121    & 85.32$\pm$0.83  & 97.75$\pm$0.70  & 75.37$\pm$8.67  & 85.48$\pm$1.44  & 97.99$\pm$0.40  & 79.21$\pm$4.34  \\
		ResNet-50       & 84.68$\pm$3.34  & 97.49$\pm$0.56  & 78.14$\pm$7.10  & 84.84$\pm$1.10  & 97.07$\pm$0.77  & 78.52$\pm$6.71  \\
		Swin-T          & \textbf{88.15$\pm$3.35} & 98.25$\pm$0.57 & \textbf{83.01$\pm$5.58} & 87.26$\pm$2.75 & \textbf{98.32$\pm$0.87} & 79.06$\pm$8.04 \\
		\bottomrule
	\end{tabular}\vspace{-20pt}
	\label{tab:internal}
\end{table}

\begin{figure}[t]
	\centering
	\subfloat[TNMammo.]{
		\includegraphics[width=0.47\linewidth]{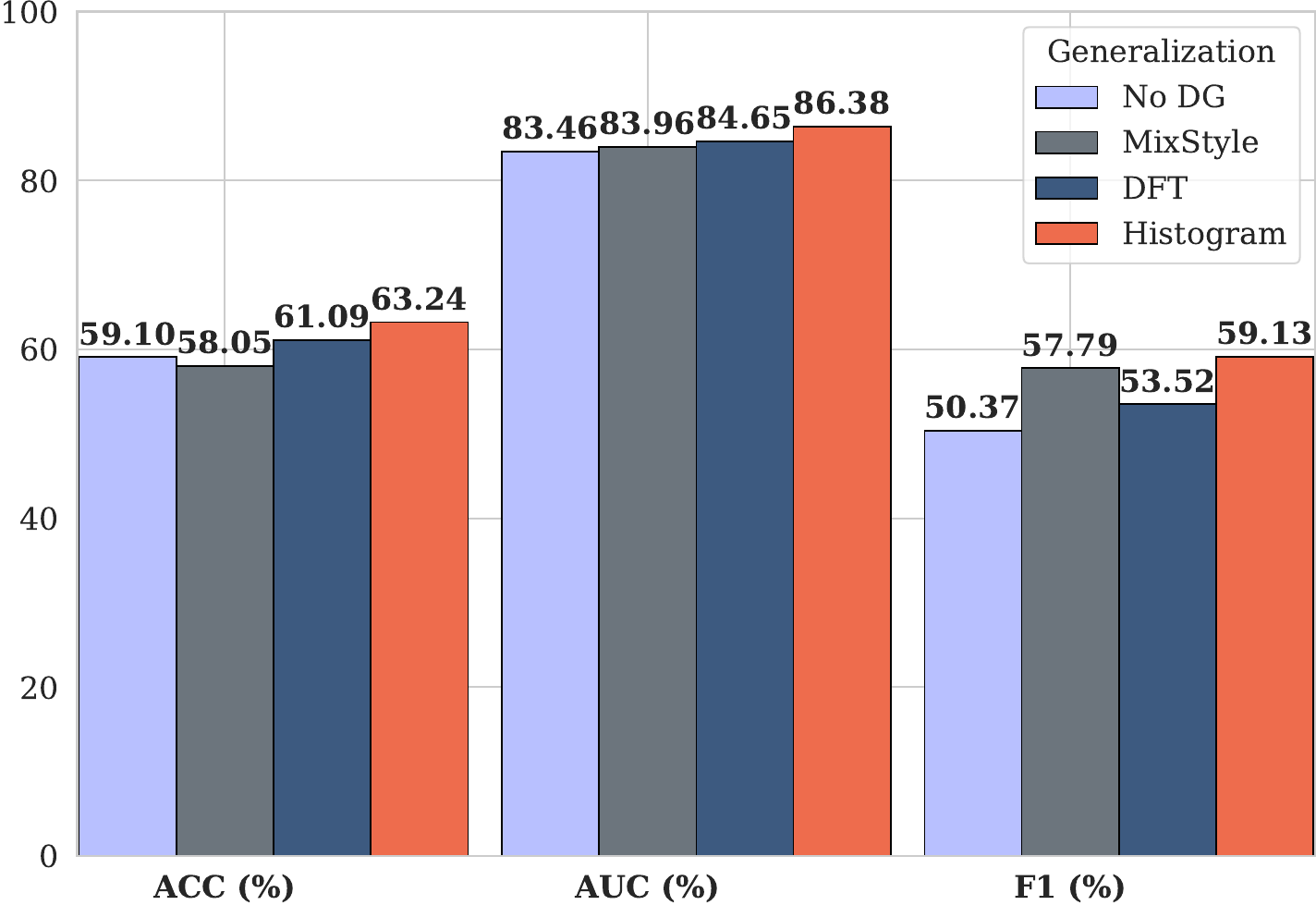}
	}
	\subfloat[LUMINA.]{
		\includegraphics[width=0.47\linewidth]{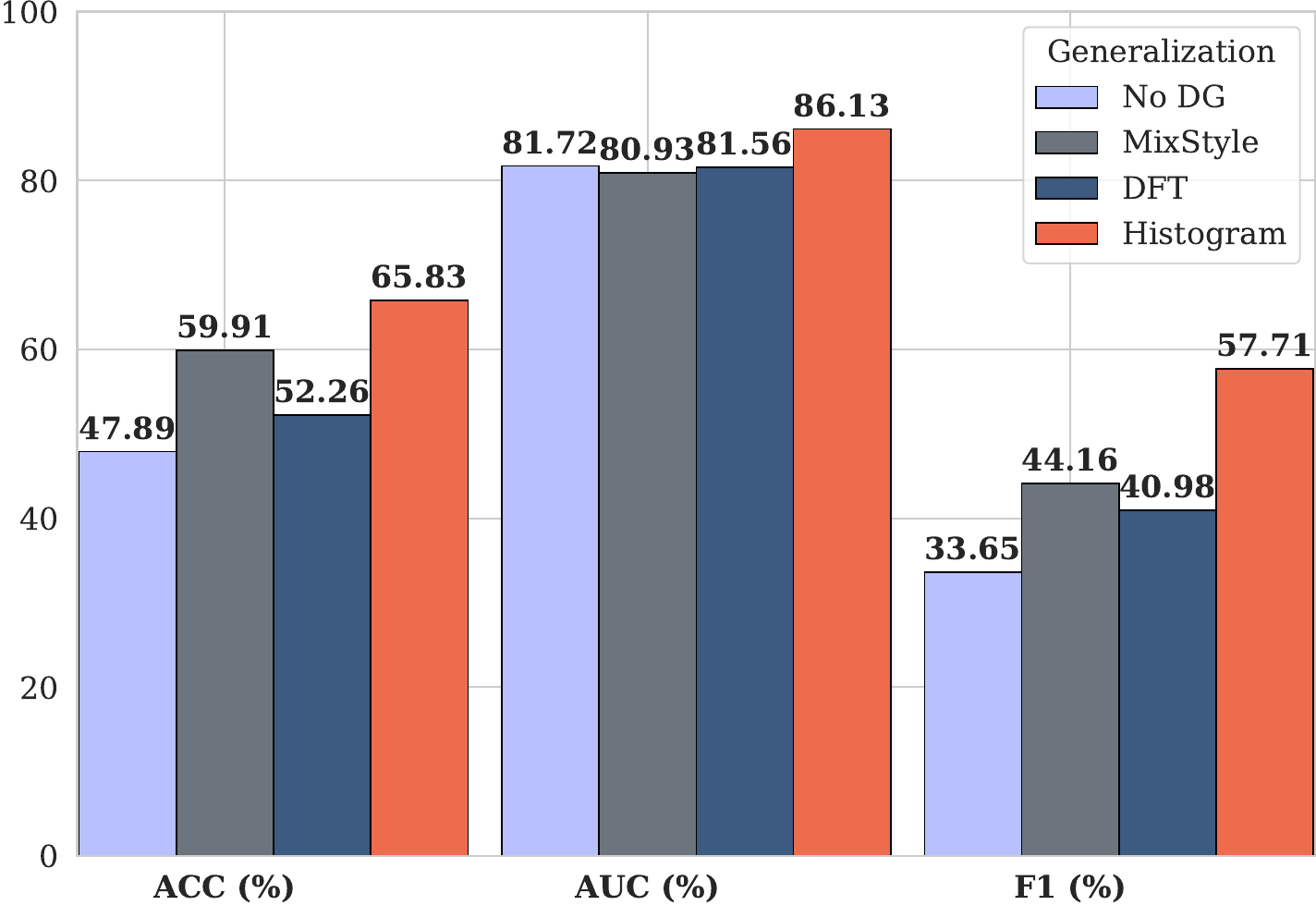}
	}\vspace{-8pt}
	\caption{\textbf{External Evaluation.} The two-view Swin-T models were trained on \textit{BreastMammo} and \textit{DenseMammo} and tested on TNMammo and LUMINA.}\vspace{-15pt}
	\label{fig:external}
\end{figure}

\noindent\textbf{External evaluation:} As shown in Fig.~\ref{fig:external}, experimental evaluation on unseen datasets demonstrates the robust generalization achieved by our approach. Our protocol successfully mitigates domain shifts by aligning pixel-space distributions across diverse vendor acquisition styles, allowing the multi-view Swin-T model to retain high diagnostic fidelity on completely novel institutional data. Specifically, our method consistently achieved top-tier performance across both external datasets. It improved the AUC from 83.46\% to 86.38\% on the TNMammo dataset and from 81.72\% to 86.13\% on the LUMINA dataset, with corresponding consistent gains in ACC and F1-score. On the contrary, alternative domain generalization methods showed limited or inconsistent efficacy. 

\begin{figure}[t]
	\centering
	\subfloat[Source.]{
		\includegraphics[width=0.18\linewidth, height=0.18\linewidth]{figure/hist/44LCC.png}
	}
	\subfloat[Reference.]{
		\includegraphics[width=0.18\linewidth, height=0.18\linewidth]{figure/hist/3RCC.png}
	}
	\subfloat[Vanilla DFT.]{
		\includegraphics[width=0.18\linewidth, height=0.18\linewidth]{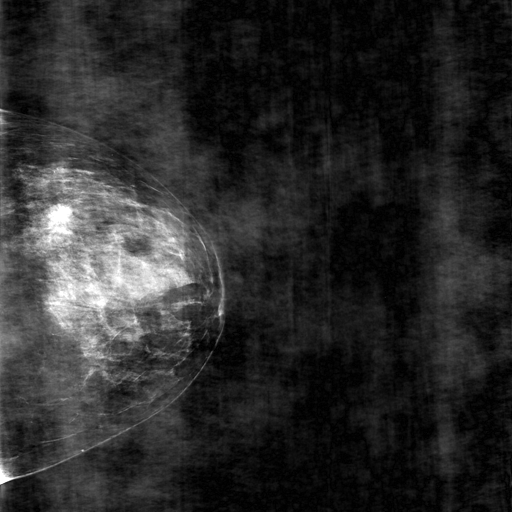}
	}
	\subfloat[DFT+Mask.]{
		\includegraphics[width=0.18\linewidth, height=0.18\linewidth]{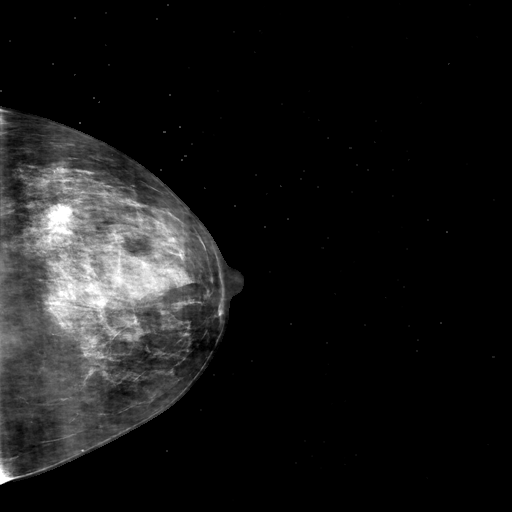}
	}
	\subfloat[Histogram.]{
		\includegraphics[width=0.18\linewidth, height=0.18\linewidth]{figure/hist/4.png}
	}\vspace{-8pt}
	\caption{\textbf{Comparison of Intensity Alignment Frameworks.} 
		The objective is to adapt the acquisition style of the source image (a) to match the reference target (b). 
		(c) Vanilla DFT interpolation displays prominent artifact bleeding and structural distortion across the entire image frame. (d) Even with a post-processing spatial mask to clean the background, frequency leakage remains irreversibly embedded within the tissue region. (e) Our proposed foreground-only histogram matching protocol completely prevents background interference.}\vspace{-10pt}
	\label{fig:dftvshistogram}
\end{figure}

\noindent\textbf{Discussion 1: Why data expansion surpasses feature-space manipulation?} 
While MixStyle introduces regularization by implicitly blending feature-map statistics within individual mini-batches, this abstract manipulation lacks explicit anatomical grounding and can inadvertently corrupt the delicate, localized texture patterns critical for distinguishing tissue densities. 
In contrast, our foreground-only histogram matching protocol demonstrates that explicitly expanding the training manifold via high-fidelity, pixel-space synthetic images is far more efficient and structurally reliable. By directly generating a continuous spectrum of tissue-specific intensity profiles, we expose the network to diverse vendor characteristics while strictly preserving the underlying anatomical structures. 
This data expansion allows models to naturally learn robust, domain-invariant representations from intact clinical morphology, resulting in superior feature scaling when deployed to entirely unseen external institutions.

\noindent\textbf{Discussion 2: Why foreground histogram matching outperforms DFT on mammograms?} As shown in Fig.~\ref{fig:dftvshistogram}, the Fourier-based domain generalization takes the massive and uninformative black background into the low-frequency style calculation. By contrast, our model-agnostic histogram matching protocol operates strictly on dynamically isolated foreground pixels. By completely ignoring the background from the outset, it ensures that intensity alignment is purely driven by actual breast tissue density distributions. This continuous blending effectively harmonizes vendor-specific contrast variances without introducing frequency leakage or artificial edge discontinuities, directly leading to the superior generalization performance shown in Fig.~\ref{fig:external}.

\section{Conclusion}
In this work, we presented a robust, anatomically grounded framework for multi-view mammography analysis designed to overcome the pervasive challenges of vendor-induced domain shift. To support the research community in developing reliable clinical models, we introduced and benchmarked two distinct multi-view FFDM datasets \textit{BreastMammo} and \textit{DenseMammo}. By integrating a novel foreground-only histogram matching protocol with a two-view Swin-T architecture, we successfully reduced the mammography domain shift on the external datasets. Our external validation on the unseen TNMammo and LUMINA datasets demonstrates that our localized, pixel-space alignment strategy consistently improves model generalization, significantly outperforming traditional frequency-space interpolation and feature-level statistical blending paradigms. These results underscore the critical importance of tissue-specific intensity alignment for building scalable, site-agnostic deep learning frameworks capable of managing the masking effect in diverse clinical screening environments.

\section*{Acknowledgment}
This research was partially funded by NIH: R01-HL171376.

\section*{Data and Code Availability}
\noindent\textbf{BreastMammo:} \url{https://osf.io/n4yr2/} \quad
\textbf{DenseMammo:} \url{https://osf.io/4azcr/} \quad
\textbf{Source code:} \url{https://github.com/NUBagciLab/BreastMammo/}

%
%
%
\bibliographystyle{splncs04}
\bibliography{main}

@String(ICASSP=	{ICASSP})

@article{lee2017curated,
  title={A curated mammography data set for use in computer-aided detection and diagnosis research},
  author={Lee, Rebecca Sawyer and Gimenez, Francisco and Hoogi, Assaf and Miyake, Kanae Kawai and Gorovoy, Mia and Rubin, Daniel L},
  journal={Scientific data},
  volume={4},
  number={1},
  pages={1--9},
  year={2017},
  publisher={Nature Publishing Group}
}

@article{moreira2012inbreast,
  title={Inbreast: toward a full-field digital mammographic database},
  author={Moreira, In{\^e}s C and Amaral, Igor and Domingues, In{\^e}s and Cardoso, Ant{\'o}nio and Cardoso, Maria Joao and Cardoso, Jaime S},
  journal={Academic radiology},
  volume={19},
  number={2},
  pages={236--248},
  year={2012},
  publisher={Elsevier}
}

@article{nguyen2023vindr,
  title={VinDr-Mammo: A large-scale benchmark dataset for computer-aided diagnosis in full-field digital mammography},
  author={Nguyen, Hieu T and Nguyen, Ha Q and Pham, Hieu H and Lam, Khanh and Le, Linh T and Dao, Minh and Vu, Van},
  journal={Scientific Data},
  volume={10},
  number={1},
  pages={277},
  year={2023},
  publisher={Nature Publishing Group UK London}
}

@article{cai2023online,
  title={An online mammography database with biopsy confirmed types},
  author={Cai, Hongmin and Wang, Jinhua and Dan, Tingting and Li, Jiao and Fan, Zhihao and Yi, Weiting and Cui, Chunyan and Jiang, Xinhua and Li, Li},
  journal={Scientific Data},
  volume={10},
  number={1},
  pages={123},
  year={2023},
  publisher={Nature Publishing Group UK London}
}

@article{shen2007image,
  title={Image registration by local histogram matching},
  author={Shen, Dinggang},
  journal={Pattern Recognition},
  volume={40},
  number={4},
  pages={1161--1172},
  year={2007},
  publisher={Elsevier}
}

@inproceedings{shapira2013multiple,
  title={Multiple histogram matching},
  author={Shapira, Dori and Avidan, Shai and Hel-Or, Yacov},
  booktitle={2013 IEEE international conference on image processing},
  pages={2269--2273},
  year={2013},
  organization={IEEE}
}

@article{rolland2000fast,
  title={Fast algorithms for histogram matching: Application to texture synthesis},
  author={Rolland, Jannick P and Vo, V and Bloss, B and Abbey, Craig K},
  journal={Journal of Electronic Imaging},
  volume={9},
  number={1},
  pages={39--45},
  year={2000},
  publisher={SPIE}
}

@inproceedings{tu2013histogram,
  title={Histogram equalization and image feature matching},
  author={Tu, Liangping and Dong, Changqing},
  booktitle={2013 6th International Congress on Image and Signal Processing (CISP)},
  volume={1},
  pages={443--447},
  year={2013},
  organization={IEEE}
}

@inproceedings{liu2021feddg,
  title={Feddg: Federated domain generalization on medical image segmentation via episodic learning in continuous frequency space},
  author={Liu, Quande and Chen, Cheng and Qin, Jing and Dou, Qi and Heng, Pheng-Ann},
  booktitle={Proceedings of the IEEE/CVF conference on computer vision and pattern recognition},
  pages={1013--1023},
  year={2021}
}

@article{lukasiewicz2021breast,
  title={Breast cancer—epidemiology, risk factors, classification, prognostic markers, and current treatment strategies—an updated review},
  author={{\L}ukasiewicz, Sergiusz and Czeczelewski, Marcin and Forma, Alicja and Baj, Jacek and Sitarz, Robert and Stanis{\l}awek, Andrzej},
  journal={Cancers},
  volume={13},
  number={17},
  pages={4287},
  year={2021},
  publisher={MDPI}
}

@article{wilkinson2022understanding,
  title={Understanding breast cancer as a global health concern},
  author={Wilkinson, Louise and Gathani, Toral},
  journal={The British journal of radiology},
  volume={95},
  number={1130},
  pages={20211033},
  year={2022},
  publisher={The British Institute of Radiology.}
}

@article{ries2008seer,
  title={SEER cancer statistics review, 1975--2005},
  author={Ries, LAG and Melbert, D and Krapcho, M and Stinchcomb, DG and Howlader, N and Horner, MJ and Mariotto, A and Miller, BA and Feuer, EJ and Altekruse, SF and others},
  journal={Bethesda, MD: National Cancer Institute},
  volume={2999},
  year={2008}
}

@article{mann2022breast,
  title={Breast cancer screening in women with extremely dense breasts recommendations of the European Society of Breast Imaging (EUSOBI)},
  author={Mann, Ritse M and Athanasiou, Alexandra and Baltzer, Pascal AT and Camps-Herrero, Julia and Clauser, Paola and Fallenberg, Eva M and Forrai, Gabor and Fuchsj{\"a}ger, Michael H and Helbich, Thomas H and Killburn-Toppin, Fleur and others},
  journal={European radiology},
  volume={32},
  number={6},
  pages={4036--4045},
  year={2022},
  publisher={Springer}
}

@inproceedings{huang2017densely,
  title={Densely connected convolutional networks},
  author={Huang, Gao and Liu, Zhuang and Van Der Maaten, Laurens and Weinberger, Kilian Q},
  booktitle={Proceedings of the IEEE conference on computer vision and pattern recognition},
  pages={4700--4708},
  year={2017}
}

@article{loshchilov2017decoupled,
  title={Decoupled weight decay regularization},
  author={Loshchilov, Ilya and Hutter, Frank},
  journal={arXiv preprint arXiv:1711.05101},
  year={2017}
}

@inproceedings{deng2009imagenet,
  title={Imagenet: A large-scale hierarchical image database},
  author={Deng, Jia and Dong, Wei and Socher, Richard and Li, Li-Jia and Li, Kai and Fei-Fei, Li},
  booktitle={2009 IEEE conference on computer vision and pattern recognition},
  pages={248--255},
  year={2009},
  organization={Ieee}
}

@inproceedings{pan2024domain,
  title={Domain generalization with fourier transform and soft thresholding},
  author={Pan, Hongyi and Wang, Bin and Zhang, Zheyuan and Zhu, Xin and Jha, Debesh and Cetin, Ahmet Enis and Spampinato, Concetto and Bagci, Ulas},
  booktitle={ICASSP 2024-2024 IEEE International Conference on Acoustics, Speech and Signal Processing (ICASSP)},
  pages={2106--2110},
  year={2024},
  organization={IEEE}
}

@inproceedings{pan2025frequency,
  title={Frequency-based federated domain generalization for polyp segmentation},
  author={Pan, Hongyi and Jha, Debesh and Biswas, Koushik and Bagci, Ulas},
  booktitle={ICASSP 2025-2025 IEEE International Conference on Acoustics, Speech and Signal Processing (ICASSP)},
  pages={1--5},
  year={2025},
  organization={IEEE}
}

@inproceedings{suckling1994mammographic,
  title={The mammographic images analysis society digital mammogram database},
  author={Suckling, John},
  booktitle={Exerpta Medica. International Congress Series, 1994},
  volume={1069},
  pages={375--378},
  year={1994}
}

@article{carr2022rsna,
  title={RSNA screening mammography breast cancer detection},
  author={Carr, C and Kitamura, Felipe and Partridge, G and Kalpathy-Cramer, J and Mongan, J and Andriole, K and Lavender, Vazirabad M and Riopel, M and Ball, R and Dane, S and others},
  journal={Kaggle},
  year={2022}
}

@article{khaled2022categorized,
  title={Categorized contrast enhanced mammography dataset for diagnostic and artificial intelligence research},
  author={Khaled, Rana and Helal, Maha and Alfarghaly, Omar and Mokhtar, Omnia and Elkorany, Abeer and El Kassas, Hebatalla and Fahmy, Aly},
  journal={Scientific data},
  volume={9},
  number={1},
  pages={122},
  year={2022},
  publisher={Nature Publishing Group UK London}
}

@article{alsolami2021king,
  title={King abdulaziz university breast cancer mammogram dataset (KAU-BCMD)},
  author={Alsolami, Asmaa S and Shalash, Wafaa and Alsaggaf, Wafaa and Ashoor, Sawsan and Refaat, Haneen and Elmogy, Mohammed},
  journal={Data},
  volume={6},
  number={11},
  pages={111},
  year={2021},
  publisher={MDPI}
}

@inproceedings{tan2019efficientnet,
  title={Efficientnet: Rethinking model scaling for convolutional neural networks},
  author={Tan, Mingxing and Le, Quoc},
  booktitle={International conference on machine learning},
  pages={6105--6114},
  year={2019},
  organization={PMLR}
}

@inproceedings{he2016deep,
  title={Deep residual learning for image recognition},
  author={He, Kaiming and Zhang, Xiangyu and Ren, Shaoqing and Sun, Jian},
  booktitle={Proceedings of the IEEE conference on computer vision and pattern recognition},
  pages={770--778},
  year={2016}
}

@inproceedings{liu2021swin,
  title={Swin transformer: Hierarchical vision transformer using shifted windows},
  author={Liu, Ze and Lin, Yutong and Cao, Yue and Hu, Han and Wei, Yixuan and Zhang, Zheng and Lin, Stephen and Guo, Baining},
  booktitle={Proceedings of the IEEE/CVF international conference on computer vision},
  pages={10012--10022},
  year={2021}
}

@article{nguyen2025tn,
  title={TN-Mammo: A Multi-view Mammography Dataset for Breast Density Classification},
  author={Nguyen, Binh and Le, Cat and Vu, Loc and Nguyen, Quynh and Pham, Ha-Hieu and Vu, Phuong Anh and Huynh, Thuan and Goldberger, A and Amaral, L and Glass, L and others},
  year={2025}
}

@inproceedings{xu2021fourier,
  title={A fourier-based framework for domain generalization},
  author={Xu, Qinwei and Zhang, Ruipeng and Zhang, Ya and Wang, Yanfeng and Tian, Qi},
  booktitle={Proceedings of the IEEE/CVF conference on computer vision and pattern recognition},
  pages={14383--14392},
  year={2021}
}

@inproceedings{pan2026lumina,
  title={LUMINA: A Multi-Vendor Mammography Benchmark with Energy Harmonization Protocol},
  author={Pan, Hongyi and Durak, Gorkem and Aktas, Halil Ertugrul and Bejar, Andrea M and Tutun, Baver and Uysal, Emre and Bulbul, Ezgi and Dogan, Mehmet Fatih and Erok, Berrin and Yildirim, Berna Akkus and others},
  booktitle={Proceedings of the IEEE/CVF Conference on Computer Vision and Pattern Recognition},
  pages={35301--35310},
  year={2026}
}

@article{zafari2026mammoclean,
  title={MammoClean: Toward Reproducible and Bias-Aware AI in Mammography through Dataset Harmonization},
  author={Zafari, Yalda and Pan, Hongyi and Durak, Gorkem and Bagci, Ulas and Rashed, Essam A and Mabrok, Mohamed},
  journal={IEEE Access},
  year={2026},
  publisher={IEEE}
}

@article{zhou2021domain,
  title={Domain generalization with mixstyle},
  author={Zhou, Kaiyang and Yang, Yongxin and Qiao, Yu and Xiang, Tao},
  journal={arXiv preprint arXiv:2104.02008},
  year={2021}
}

@article{paszke2019pytorch,
  title={Pytorch: An imperative style, high-performance deep learning library},
  author={Paszke, Adam and Gross, Sam and Massa, Francisco and Lerer, Adam and Bradbury, James and Chanan, Gregory and Killeen, Trevor and Lin, Zeming and Gimelshein, Natalia and Antiga, Luca and others},
  journal={Advances in neural information processing systems},
  volume={32},
  year={2019}
}
\end{document}